\documentclass[11pt,leqno]{mf}
\usepackage{graphicx} \usepackage{amssymb}
\usepackage{amsmath}
\newcommand{\hsc}{\hspace*{-1em}\sc }

\title{ A CONVEX STOCHASTIC OPTIMIZATION PROBLEM ARISING FROM
PORTFOLIO SELECTION\footnotetext{Supported by the RGC Earmarked Grants
CUHK 4175/03E, CUHK418605, and Croucher Senior Research Fellowship.}
\footnotetext{Address corresponding to Xun Yu Zhou, Department of Systems Engineering and Engineering Management, The Chinese
University of Hong Kong, Shatin, Hong Kong. Tel.: 852-2609-8320, fax: 852-2603-5505}
}

\author{ {\sc Hanqing Jin, Zuo Quan Xu and Xun Yu Zhou
}\\
\it The Chinese University of Hong Kong} 

\newcommand{\dpm}{{\prime\hspace{-0.03cm}\prime}}

\newtheorem{theo}{\quad \sc Theorem}[section]
\newtheorem{lemma}{\quad \sc Lemma}[section]
\newtheorem{prop}{\quad \sc Proposition}[section]
\newtheorem{coro}{\quad \sc Corollary}[section]

\newtheorem{remark}{\quad \sc Remark}[section]
\newtheorem{ex}{\quad \sc Example}[section]

\newcommand{\eof}{\hfill $\Box$ \par}

\newcommand{\pf}{{\it Proof: }}

\newcommand{\R}{{\hbox{I{\kern -0.22em}R}}}
\newcommand{\N}{{\hbox{I{\kern -0.22em}N}}}
\newcommand{\sublim}{\liminf}
\newcommand{\suplim}{\limsup}
\newcommand{\id}{{\mathbf 1}}
\newcommand{\einf}{{\rm essinf \;}}
\numberwithin{equation}{section}

\begin{document}
\date{}
\maketitle

\begin{abstract}
A continuous-time financial portfolio selection model with expected
utility maximization typically boils down to
solving a (static) convex stochastic optimization problem in terms of
the terminal
wealth, with a budget constraint.
In literature the latter is solved by assuming {\it a priori}
that the problem is well-posed (i.e., the supremum value is finite) and
a Lagrange multiplier exists (and as a consequence the optimal solution
is attainable).
In this paper it is first shown, via various counter-examples, neither
of these two assumptions needs to hold, and an optimal solution
does not necessarily exist. These anomalies in turn have important
interpretations in and impacts on the portfolio selection modeling and
solutions.
Relations among the non-existence of the Lagrange multiplier, the ill-posedness
of the problem, and the non-attainability of an optimal solution are
then investigated. Finally, explicit and
easily verifiable conditions are derived
which lead to finding the unique optimal solution.

\smallskip


\end{abstract}

\section{Introduction}
Given a probability space $(\Omega,{\mathcal F},P)$,
consider the following constrained stochastic optimization problem
\begin{equation}\label{maxut}
\begin{array}{ll}
{\rm Maximize}& Eu(X)\\
\mbox{\rm subject to }& E[X\xi]=a,\;\; X\ge 0 \mbox{ is a random variable},
\end{array}
\end{equation}
where $a>0$ is a parameter,
$\xi>0$ a given scalar-valued random variable,
$u(\cdot)$: $\R^{+}\mapsto \R^{+}$ a twice differentiable,
strictly increasing, strictly concave function with $u(0)=0, 
u'(0+)=+\infty, u'(+\infty)=0$.
Define $V(a)=\sup\nolimits_{E[X\xi]=a, X\ge 0 \mbox{ is a r.v.}}
Eu(X)$.

It is well known that many continuous-time financial
portfolio selection problems with
expected utility maximization
boil down to solving problem (\ref{maxut}). In the context of a
portfolio model,
$u(\cdot)$ is the utility function (all the assumed properties on
$u(\cdot)$ have economic interpretations), $\xi$ is the so-called pricing kernel or
state price density, $a$ is the initial wealth (hence the first constraint is the budget constraint),
and $X$ is the terminal wealth to be determined. Once an optimal $X^*$
to (\ref{maxut}) is found,
the portfolio replicating $X^*$ is the optimal portfolio for the original
dynamic portfolio choice problem, if the market is complete.
For details see, e.g., Cvitanic and Karatzas (1992), Karatzas (1997),
Karatzas and Shreve (1998), Korn (1997).


In literature (\ref{maxut}) is
usually solved by the Lagrange method, which is summarized in the following theorem.

\begin{theo}\label{generallag}
  If (\ref{maxut}) admits an optimal solution $X^*$ whose objective value
is finite, 
  then there exists
  $\lambda>0$ such that $X^*=(u')^{-1}(\lambda \xi)$.
Conversely, if $E[(u')^{-1}(\lambda \xi)\xi]=a<+\infty$
  and $E[u\left((u')^{-1}(\lambda \xi)\right)]<+\infty$,
  then $X^*=(u')^{-1}(\lambda \xi)$ is optimal for (\ref{maxut})
  with parameter $a$.
\end{theo}

This theorem  provides an efficient scheme to find the optimal solution
for  Problem (\ref{maxut}): For any $a>0$, solve the Lagrange equation
$E[(u')^{-1}(\lambda \xi)\xi]=a$ -- {\it if one could} --
 to determine a Lagrange multiplier $\lambda$,
and then $X^*=(u')^{-1}(\lambda \xi)$ is the optimal (automatically unique
as the utility function is strictly concave) solution for (\ref{maxut}),
{\it if} $Eu(X^*)$ is finite.

However, there are many issues about Problem (\ref{maxut}) that are left
untouched by the preceding theorem/scheme.
To elaborate, in general there are the following progressive issues related to an optimization problem
such as (\ref{maxut}):

\begin{itemize}
\item {\it Feasibility}: whether there is at least one solution satisfying
all the constraints involved. For (\ref{maxut}), since $X=a/\xi$ is a feasible solution,
the feasibility is not an issue.\footnote{Feasibility
could be by itself an interesting problem if more complex constraints
are involved. See Section 3 of Bielecki et. al (2005) for an example.}
\item {\it Well-posedness}: whether the supremum value of the problem
with a non-empty feasible set is
finite (in which case the problem is called well-posed) or $+\infty$ (ill-posed).
An ill-posed problem is a mis-formulated one: the trade-off is not set right so
one could always push the objective value to be arbitrarily high.\footnote{Again, well-posedness is an important, sometimes {\it very}
difficult, problem in its
own right; see Jin and Zhou (2006) for a behavioral portfolio selection model where
the well-posedness becomes an eminent issue. Also see Korn and Kraft (2004)
for more ill-posed examples.}

\item {\it Attainability}: whether a well-posed problem admits
an optimal solution. It may or may not.
\item {\it Uniqueness}: whether an attainable problem has a unique optimal
solution. It is not an issue for (\ref{maxut}), since uniqueness holds automatically
due to the strict concavity of the utility function.
\end{itemize}

Clearly, Theorem \ref{generallag} covers only the case when the problem
is well-posed and the attainability holds, by assuming {\it a priori} that
a Lagrange multiplier exists (indeed, in the context
of portfolio selection the existing work always {\it assumes} that
the Lagrange multiplier exists; see Theorem 2.2.2 in page 7 of Karatzas (1997)
and page 65 of Korn (1997)\footnote{In these references it is assumed that
$f(\lambda)=E[(u')^{-1}(\lambda \xi)\xi]<+\infty$ for any
$\lambda>0$, which is equivalent to the existence of the Lagrange multiplier
for any $a>0$; see Section 2 for details.}). Moreover, in Theorem 2.2.2 in page 7 of Karatzas (1997)
and Assumption 6.2 in page 773 of Cvitanic and Karatzas (1992),
it is assumed up front that the
underlying problem is well-posed.\footnote{Some of the references cited
here deal with models with consumptions; yet the essence of the
Lagrange method remains the same.} In this paper we will first show, through
various counter-examples, that
none of the aforementioned assumptions that have all along been
taken for granted needs to hold true.
Then, we will address the following questions:
When does the Lagrange multiplier exist?
What if it does not? What does it have to do with the well-posedness and
attainability? What are the conditions ensuring
the existence of a unique optimal solution for (\ref{maxut}) for
a given $a>0$ or for any $a>0$?

The aim of this paper is to give a thorough treatment of
(\ref{maxut}), including answers to the above questions. In particular,
Section 2 reveals
the possibility of non-existence of
the Lagrange multiplier.
Section 3 studies the implications of the non-existence of the Lagrange multiplier,
and Section 4 shows the possibility of ill-posedness even with the existence of the
Lagrange multiplier.
Finally, Section 5 presents easily verifiable conditions for
uniquely solving (\ref{maxut}).

\section{Non-Existence of Lagrange Multiplier}

It is possible that the Lagrange multiplier simply does not exist, which
will be demonstrated in this section via several examples.

First off,
define
\begin{equation}\label{flambda}
f(\lambda)=E[(u')^{-1}(\lambda \xi)\xi], \;\;\;\lambda>0.
\end{equation}
Then $f(\cdot)$ is non-increasing (notice that
$f(\cdot)$ may take value $+\infty$). The following lemma is evident
given the monotonicity of $(u')^{-1}(\cdot)$ and the monotone convergence
theorem.

\begin{lemma}\label{semicont}
If $f(\lambda_0)<+\infty$ for some $\lambda_0>0$, then $f(\cdot)$ is
continuous on $(\lambda_0, +\infty)$
and right continuous at $\lambda_0$, with $f(+\infty)=0$.
\end{lemma}

It follows from Lemma \ref{semicont} that if $f(\lambda_0)<+\infty$ for some
$\lambda_0>0$, then the Lagrange multiplier exists
for any $0<a\leqslant a_0:= E[(u')^{-1}(\lambda_0 \xi)\xi]$. In particular,
if
\begin{equation}\label{flambdafinite}
f(\lambda)<+\infty\;\;\forall \lambda>0,
\end{equation}
then the Lagrange multiplier exists
for any $a>0$. This is why
in existing literature (\ref{flambdafinite}) is usually assumed up front
(see, e.g., Karatzas (1997), p. 37, (2.2.11)
and Korn (1997), p. 65, (24)). Now, we are to show that
this assumption may not hold even for simple cases.

\begin{ex}\label{f+infty}
{\rm Take $u(x)=\sqrt{x}$, $x\geq0$, $P(\xi\le t)=1-e^{-t}$, $t\geq0$.
In this example, $u'(x)=\frac{1}{2\sqrt{x}}, (u')^{-1}(y)=(2y)^{-2}$, and
$f(\lambda)=E[(u')^{-1}(\lambda \xi)\xi]=\frac{1}{4\lambda^2}E\xi^{-1}=+\infty$
for any $\lambda>0$. Therefore
$E[(u')^{-1}(\lambda \xi)\xi]=a$ admits no solution for {any} $a>0$.
}
\end{ex}

In the above example the Lagrange multiplier does not exist for {\it any} $a>0$.
In the following examples, Lagrange multipliers
exist for {\it some} $a>0$, and do not for {\it other} $a>0$.

\begin{ex}\label{fjump}
{\rm Define $p(x)=\frac{e^x-1-x-x^2/2! -x^3/3! }{x^2}=\sum_{n=2}^{+\infty}\frac{x^n}{(n+2)!},\;
g(x)=p(\frac{1}{x}), \; h(x)=g^{-1}(x),\;x>0$. Take
$$u(x)=\left\{\begin{array}{lll}
xh(x)+\int_0^{1/h(x)}\frac{p(y)}{y^2}dy,& x>0,\\
0,&x=0,
\end{array}\right.$$
and $P(\xi\ge t)=1-e^{-1/t}$, $t>0$; or $1/\xi$ follows the exponential distribution
with parameter $1$.

In this example, $p(\cdot)$ is strictly increasing with $p(0+)=0$, $p(+\infty)=+\infty$; hence
$g(\cdot)$ is strictly decreasing with $g(0+)=+\infty, g(+\infty)=0$,
and $h(\cdot)$ is well-defined
and strictly decreasing with $h(0+)=+\infty, h(+\infty)=0$. All these functions are smooth.

For the utility function $u(\cdot)$, notice that
$\int_0^x\frac{p(y)}{y^2}dy=\int_0^x\sum_{n=0}^{\infty}\frac{y^n}{(n+4)!}dy
=\sum_{n=0}^{+\infty}\frac{x^{n+1}}{(n+4)!(n+1)}$ is well-defined for any $x>0$, and
$$\lim_{x\rightarrow 0+}xh(x)=\lim_{y\rightarrow +\infty}g(y)y
=\lim_{y\rightarrow +\infty}p(\frac{1}{y})y=0,$$
which means that $u(\cdot)$ is right-continuous at $0$. Furthermore, for any $x>0$
\begin{eqnarray*}
  u'(x)&=&h(x)+x h'(x)-
  \frac{p(1/h(x))}{1/h(x)^2}\frac{h'(x)}{h(x)^2}\\
  &=&h(x)+x h'(x)-p(1/h(x))h'(x)\\
  &=&h(x)+x h'(x)-g(h(x))h'(x)\\
  &=&h(x).
\end{eqnarray*}
Therefore $u(\cdot)$ is concave and $u'(0+)=h(0+)=+\infty, u'(+\infty)=h(+\infty)=0$.
Moreover, $u'(x)=h(x)$,
and $(u')^{-1}(y)=g(y)=\sum_{n=2}^{+\infty}\frac{1}{(n+2)!y^n}$.
On the other hand, from the distribution of $\xi$ it follows
easily that $E\xi^{-n}=n!$ for any $n\in \N$.

Now let us calculate $f(\lambda)=E[(u')^{-1}(\lambda \xi)\xi]$ for any $\lambda> 0$:
\begin{eqnarray*}
  f(\lambda)&=&E[g(\lambda \xi)\xi]\\
  &=&E[\sum_{n=2}^{+\infty}\frac{1}{(n+2)!\lambda^n}\xi^{-(n-1)}]\\
  &=&\sum_{n=2}^{+\infty}\frac{(n-1)!}{(n+2)!\lambda^n}\\
  &=&\sum_{n=2}^{+\infty}\frac{1}{(n+2)(n+1)n}\left(\frac{1}{\lambda}\right)^n.\\
\end{eqnarray*}
By the convergence of series, we know that
$f(\lambda)<+\infty$
if and only if $\lambda\ge 1$.

Define $a_1=f(1)=E[(u')^{-1}(\xi)\xi]=\sum_{n=2}^{+\infty}\frac{1}{(n+2)(n+1)n}=\frac{1}{12}$.
Then for any $0<a\le a_1$,
we can find a Lagrange multiplier $\lambda\ge 1$ such that $E[(u')^{-1}(\lambda \xi)\xi]=a$.
On the other hand, the Lagrange multiplier is non-existent when $a>a_1$.
}
\end{ex}

In the preceding examples $\xi$ is related to the exponential
distribution, whereas in applying to portfolio selection $\xi$ is
typically lognormal. The next example
shows such a case.

\begin{ex}\label{fjump2}
{\rm
Take a positive random variable $\xi$ satisfying
$0<E[\xi^{-(n-1)}]<+\infty$ $\forall n\geq 1$ and $\lim_{n\to+\infty}\frac{E[\xi^{-(n-1)}]}{E[\xi^{-n}]}=0$
(e.g., when $\xi$ is lognormal).
Define $a_{n}=\frac{1}{n^2E[\xi^{-(n-1)}]}$, $n\ge 2$, and
$p(x)=\sum_{n=2}^{+\infty}a_{n}x^n$,
$g(x)=p(\frac{1}{x})$, $h(x)=g^{-1}(x)$, $x>0$. Take
\begin{eqnarray*}
u(x)=\left\{\begin{array}{lll}
x h(x)+\int_0^{1/h(x)}\frac{p(y)}{y^2}dy,& x>0,\\
0,&x=0.
\end{array}\right.
\end{eqnarray*}

Exactly the same analysis as in Example \ref{fjump}
yields that
$u(\cdot)$ is a utility function satisfying all the required
conditions, with $u'(x)=h(x)$ and
$(u')^{-1}(x)=g(x)=\sum_{n=2}^{+\infty}a_{n}x^{-n}$.

Now, for any $\lambda> 0$,
\[  f(\lambda)=E[g(\lambda \xi)\xi]
=E[\sum_{n=2}^{+\infty}a_{n}\lambda^{-n}\xi^{-(n-1)}]
=\sum_{n=2}^{+\infty}\frac{1}{n^2\lambda^n}.
\]
Hence $f(\lambda)<+\infty$
if and only if $\lambda\ge1$.
As a result, the Lagrange multiplier exists if and only if
$0<a\le a_1$,
where
$a_1=f(1)=E[(u')^{-1}(\xi)\xi]=\sum_{n=2}^{+\infty}\frac{1}{n^2}=\frac{\pi^2-6}{6}$.
}
\end{ex}

\section{Implication of Non-Existence of Lagrange Multiplier}

So, if the Lagrange multiplier does not exist, what can we say about
the underlying optimization problem (\ref{maxut})? Theorem \ref{generallag}
implies that the non-existence of the Lagrange multiplier is an indication of
either the ill-posedness or the non-attainability of
(\ref{maxut}).
In this section we
elaborate on this.


\begin{theo}\label{lambda0infty}
  If $E[(u')^{-1}(\lambda \xi)\xi]=+\infty$ for any $\lambda>0$, then
  $V(a)=+\infty$ for any $a>0$.
\end{theo}
\pf
Fix $\lambda_0>0$ and $a>0$. Since $E[(u')^{-1}(\lambda_0\xi)\xi]=+\infty$,
one can find a set $A\in\mathcal{F}$ such that
$E[(u')^{-1}(\lambda_0\xi)\xi\id_A]\in (a, +\infty)$.
Define $h(\lambda)=E[(u')^{-1}(\lambda\xi)\xi\id_A]$, $\lambda\in [\lambda_0, +\infty)$. Then $h(\cdot)$
is non-increasing and continuous on $[\lambda_0, +\infty)$ with $h(+\infty)=0$;
hence there exists $\lambda_1>\lambda_0$ such that $h(\lambda_1)=a$.

Denote $X_1=(u')^{-1}(\lambda_1\xi)\id_{A}$, 
which is a feasible solution for Problem (\ref{maxut})
with parameter $a$, and $V(a)\ge E[u(X_1)\id_{A}]\ge E[X_1 u'(X_1)\id_{A}]=
E[(u')^{-1}(\lambda_1\xi)\lambda_1\xi\id_{A}]=\lambda_1 a>\lambda_0 a$.
(Here we have used the fact that  $u(x)\geq xu'(x)$ $\forall x>0$
owing to the concavity of
$u(\cdot)$ and that $u(0)=0$.)
Since $\lambda_0>0$ is arbitrary,
we arrive at $V(a)\ge \lim_{\lambda_0\to+\infty}\lambda_0 a=+\infty$.
\eof

This theorem indicates that if the Lagrange multiplier does not exist for
{\it all} $a>0$, then (\ref{maxut}) is ill-posed for {\it all} $a>0$. Example \ref{f+infty}
exemplifies such a case.
Now, if the Lagrange multiplier does not exist for only {\it some} $a$ 
(such as in Examples \ref{fjump} and \ref{fjump2}), is it still
possible that (\ref{maxut}) is well-posed for the same $a$?
To study this, we need the following lemma.
\begin{lemma}\label{vconsis}
  $V(a)<+\infty$, $\forall a>0$
if and only if $\exists\; a>0$ such that $V(a)<+\infty$.
\end{lemma}
\pf It suffices to prove that if $V(a)<+\infty$ for some $a>0$ then
$V(b)<+\infty$ for any $b>0$.

For $b\ge a$, we have
\[\begin{array}{rl}
  V(b)=&\sup_{E[X\xi]=b, X\ge 0}Eu(X)
=\sup_{E[X\xi]=a, X\ge 0}
Eu\left(\frac{b}{a}X\right)\\
\le &\sup_{E[X\xi]=a, X\ge 0}
\frac{b}{a}Eu(X)=\frac{b}{a}V(a)<+\infty,
\end{array}
\]
where the first inequality is due to the concavity of $u(\cdot)$ and $u(0)=0$.

For any $0<b< a$,
\[ \begin{array}{rl}
 V(b)=&\sup_{E[X\xi]=b, X\ge 0}Eu(X)
=\sup_{E[X\xi]=a, X\ge 0}Eu\left(\frac{b}{a}X\right)\\
\le &\sup_{E[X\xi]=a, X\ge 0}Eu(X)=V(a)<+\infty,
\end{array}
\]
where the first inequality is due to $u(\cdot)$ being increasing.
The proof is complete.
\eof

\begin{coro}
If $V(a)<+\infty$ for some $a>0$, then there exists $a_0>0$ such that
Problem (\ref{maxut})
admits a unique optimal solution for all $0<a\le a_0$.
\end{coro}
\pf
It follows from Theorem \ref{lambda0infty}
that there exists $\lambda_0$ with
$E[(u')^{-1}(\lambda_0 \xi)\xi]<+\infty$;
consequently the Lagrange multiplier exists
for any $0<a\leqslant a_0:= E[(u')^{-1}(\lambda_0 \xi)\xi]$ by Lemma \ref{semicont}.
On the other hand, Lemma \ref{vconsis} yields that
$V(a)<+\infty$ for all $a$; hence the desired result follows by virtue of
Theorem \ref{generallag}.
\eof


Now let us continue with Example \ref{fjump2}.

\begin{ex}\label{nonattain}
{\rm
In Example \ref{fjump2},
take $\lambda=2$. We have proved that
$a_2:=E[(u')^{-1}(2\xi)\xi]<+\infty$.
Denote $X^*=(u')^{-1}(2\xi)$. Then
\begin{eqnarray*}
  Eu(X^*)&=&Eu(g(2\xi))\\
  &=&E[2\xi g(2\xi)+\int_0^{1/(2\xi)}\frac{p(y)}{y^2}dy]\\
  &=&2a_2+\sum_{n=2}^{+\infty}\frac{a_{n}}{n-1}E[(2\xi)^{-(n-1)}]\\
  &=&2a_2+\sum_{n=2}^{+\infty}\frac{2^{-(n-1)}}{n^2(n-1)}\\
  &<&+\infty.
\end{eqnarray*}
Theorem \ref{generallag} suggests that
$X^*$ is the unique optimal solution for (\ref{maxut}) with
parameter $a_2$ and, in particular,
$V(a_2)=Eu(X^*)<+\infty$.
By Lemma \ref{vconsis}, we know $V(a)<+\infty$ for any $a>0$, i.e.,
(\ref{maxut}) is well-posed for any $a>0$.

However, we have proved in Example \ref{fjump2}
that $E[(u')^{-1}(\lambda \xi)\xi]=a$
admits no solution for any $a>a_1$.
Therefore Problem (\ref{maxut}) with parameter $a>a_1$ is well-posed; yet it
admits no optimal solution (i.e., the problem is not attainable).}
\end{ex}

\section{Ill-posedness When Lagrange Multiplier Exists}

The last section demonstrated that one of the possible consequences of the
non-existence of a Lagrange multiplier is the ill-posedness of the
underlying optimization problem. This section aims to show via an example that
Problem (\ref{maxut}) may be ill-posed even if
the Lagrange multiplier {\it does} exist for {\it any} $a>0$.

\begin{ex}\label{vinfinite}
{\rm Let
$$u(x)=\left\{\begin{array}{lll}
\sqrt{x},& 0\le x\le 1,\\
1-\ln2+\ln(1+x),&x>1,
\end{array}\right.$$
and $\xi$ be a positive random variable such that $E[\ln\frac{1}{\xi}]=+\infty$.
It is easy to check that $u(\cdot)$ has all the
required properties, and
$$(u')^{-1}(x)=\left\{\begin{array}{lll}
\frac{1}{x}-1, & 0<x\le 0.5,\\
\frac{1}{4x^2},& x>0.5.
\end{array}\right.$$
Hence
$$
f(\lambda)=E[(u')^{-1}(\lambda\xi)\xi]=\frac{1}{\lambda}E[(1-\lambda\xi)\id_{\lambda\xi\le 0.5}]
+E[\frac{1}{4\lambda^2\xi}\id_{\lambda\xi>0.5}]\le \frac{3}{2\lambda}<+\infty\;\;\forall \lambda>0.
$$
As a result, the Lagrange multiplier exists
for any $a>0$.
However, for any $\lambda>0$,
$$
E[u((u')^{-1}(\lambda\xi))]=E[(1-\ln2-\ln(\lambda\xi))\id_{\lambda\xi\le 0.5}]
+E[\frac{1}{2\lambda\xi}\id_{\lambda\xi>0.5}]
\ge E[\ln\frac{1}{\xi}\id_{\lambda\xi\le 0.5}]-\ln(\lambda)=+\infty.
$$}
\end{ex}

\begin{remark}
{\rm In existing literature it is usually assumed, either explicitly (see,
e.g., Karatzas (1997), p. 37, (2.2.13)) or implicitly, that the problem is
well-posed for all $a$. The preceding example proves that the well-posedness
is not guaranteed even when the Lagrange multiplier exists.}
\end{remark}

\section{Optimal Solution}

Having discussed on the ill-posedness and non-attainability, we
are now in a position to study the optimal solution of (\ref{maxut}).
The problems with
Theorem \ref{generallag} are two-fold. On one hand,
the required conditions that
the Lagrange equation
$E[(u')^{-1}(\lambda \xi)\xi]=a$ admits a positive solution
and that $E[u\left((u')^{-1}(\lambda \xi)\right)]<+\infty$
do
not necessarily hold (as already demonstrated), and on the other hand
even if the conditions do hold, they are
implicit and/or hard to verify.
In this section, we will present conditions that are
explicit and easy to use.

Recall that $f(\lambda)=E[(u')^{-1}(\lambda \xi)\xi], \;\;\lambda>0$.
If $f(\lambda)=+\infty$ for any $\lambda>0$, then
it follows from Theorem \ref{lambda0infty} that $V(a)=+\infty$
for any $a>0$, which is a pathological case. Hence
we assume that there exists a $\lambda>0$ such that $f(\lambda)<+\infty$. Denote
$\lambda_0=\inf\{\lambda>0: f(\lambda)<+\infty\}<+\infty$
and $a_0=f(\lambda_0+)$ (notice that $a_0=+\infty$ is possible, and $a_0=f(\lambda_0)$ when $\lambda_0>0$).

\begin{prop}\label{gebeexist}
  Suppose $\lambda_0<+\infty$. We have the following conclusions.
\begin{itemize}
  \item[{\rm (i)}] If $a_0<+\infty$, then Problem (\ref{maxut})
with parameter $a>0$ admits a unique
optimal solution if and only if
  $E[u((u')^{-1}(\lambda_0\xi))]<+\infty$ and $a\le a_0$.
  \item[{\rm (ii)}]  If $a_0=+\infty$, then Problem (\ref{maxut})
  admits a unique optimal solution for any $a>0$ if and only if
   $E[u((u')^{-1}(\xi))]<+\infty$.
\end{itemize}
\end{prop}

\pf
(i) is clear in view of Theorem \ref{generallag} and Lemma \ref{vconsis}.
To prove (ii), if $a_0=+\infty$,
by Lemma
\ref{semicont}, $f(\cdot)$ is continuous on $(\lambda_0,+\infty)$ with
$f(\lambda_0+)=+\infty$ and $f(+\infty)=0$; hence the Lagrange multiplier
exists for any $a>0$. Now, if $E[u((u')^{-1}(\xi))]<+\infty$,
using $u(x)\geq xu'(x)$ with $x=(u')^{-1}(\xi)$, we have
\[ +\infty>E[u((u')^{-1}(\xi))]\geq E[(u')^{-1}(\xi)\xi]=:a_1.\]
It follows from Theorem \ref{generallag} that $V(a_1)=E[u((u')^{-1}(\xi))]<+\infty$.
Lemma \ref{vconsis} further yields $V(a)<+\infty$, $\forall a>0$. The desired result is now
a consequence of Theorem \ref{generallag}.
\eof

Now we derive some sufficient conditions, explicit in terms of
$u(\cdot)$ or $\xi$, for the existence of a unique optimal solution to
(\ref{maxut}). First we have the following simple case.

\begin{theo}\label{einf0}
If $\varepsilon=\einf\xi>0$, then Problem (\ref{maxut}) admits a unique optimal solution
for any $a>0$.
\end{theo}
\pf Given $a>0$. For any feasible solution $X$ of Problem (\ref{maxut}),
$$Eu(X)\le u(EX)\le u(\frac{E[X\xi]}{\varepsilon})=u(\frac{a}{\varepsilon}).$$
Therefore $V(a)<+\infty$.

Meanwhile, for any $\lambda>0$,
$$f(\lambda)=E[(u')^{-1}(\lambda \xi)\xi]\le \frac{1}{\lambda}E[u((u')^{-1}(\lambda\xi))]
\le \frac{1}{\lambda}u((u')^{-1}(\lambda\varepsilon))<+\infty.$$
This proves the existence of the Lagrange multiplier $\lambda>0$
for any $a>0$.
By Theorem \ref{generallag}, $X_\lambda=(u')^{-1}(\lambda \xi)$ is the
unique optimal solution for
(\ref{maxut}).
\eof

Let us make some preparations for our main result.

Define $R(x)=-\frac{xu^\dpm(x)}{u'(x)}\ge 0$ as the {\it Arrow--Pratt index
of risk aversion} of the utility function $u(\cdot)$.

\begin{lemma}\label{lbra}
If $\sublim_{x\rightarrow +\infty} R(x)>0$,
then $\suplim_{x\rightarrow +\infty}\frac{u'(kx)}{u'(x)}<1$
for any $k>1$.
\end{lemma}
\pf Because $\sublim_{x\rightarrow +\infty} R(x)>0$,
there exist $M>0$, $K>0$, such that $R(x)\ge K$ for any $x\ge M$.
For any $x\ge M$, $k>1$,
\begin{eqnarray*}
  \frac{u'(kx)}{u'(x)}-1&=& \frac{u'(kx)-u'(x)}{u'(x)}\\
  &=& \frac{\int_x^{kx}u^\dpm(y)dy}{u'(x)}\\
  &=& -\frac{\int_x^{kx}R(y)u'(y)/ydy}{u'(x)}\\
  &\le&-\frac{\int_x^{kx}R(y)u'(kx)/ydy}{u'(x)}\\
  &=&-\frac{u'(kx)}{u'(x)}\int_x^{kx}R(y)/ydy\\
  &\le&-\frac{u'(kx)}{u'(x)}K\int_x^{kx}1/ydy\\
  &=&-\frac{u'(kx)}{u'(x)}K\ln{k}.
\end{eqnarray*}
Therefore $\frac{u'(kx)}{u'(x)}\le \frac{1}{1+K\ln{k}}$ which implies
$\suplim_{x\rightarrow +\infty}\frac{u'(kx)}{u'(x)} \le \frac{1}{1+K\ln{k}}<1$.
\eof

\begin{lemma}\label{u'inv}
$\suplim_{x\rightarrow 0+}\frac{(u')^{-1}(\lambda x)}{(u')^{-1}(x)}<+\infty$ for any $0<\lambda<1$
if and only if $\suplim_{x\rightarrow +\infty}\frac{u'(kx)}{u'(x)}<1$ for any $k>1$.
\end{lemma}
\pf
We first claim that
$\suplim_{x\rightarrow 0+}\frac{(u')^{-1}(\lambda x)}{(u')^{-1}(x)}<+\infty$
for any $0<\lambda<1$ if and only if
$\exists\; 0<\bar\lambda<1$ such that
$\suplim_{x\rightarrow 0+}\frac{(u')^{-1}(\bar\lambda x)}{(u')^{-1}(x)}<+\infty$.

To prove this claim, suppose
$\suplim_{x\rightarrow 0+}\frac{(u')^{-1}(\bar\lambda x)}{(u')^{-1}(x)}<+\infty$
for some $0<\bar\lambda<1$. Then
\begin{eqnarray*}
&&\suplim_{x\rightarrow 0+}\frac{(u')^{-1}(\bar\lambda^2 x)}{(u')^{-1}(x)}\\
&=&\suplim_{x\rightarrow 0+}\frac{(u')^{-1}(\bar\lambda^2 x)}{(u')^{-1}(\bar\lambda x)}
         \frac{(u')^{-1}(\bar\lambda x)}{(u')^{-1}(x)}\\
&\le&\suplim_{x\rightarrow 0+}\frac{(u')^{-1}(\bar\lambda^2 x)}{(u')^{-1}(\bar\lambda x)}
     \suplim_{x\rightarrow 0+}\frac{(u')^{-1}(\bar\lambda x)}{(u')^{-1}(x)}\\
&<&+\infty.
\end{eqnarray*}
From induction it follows
$\suplim_{x\rightarrow 0+}\frac{(u')^{-1}(\bar\lambda^nx)}{(u')^{-1}(x)}<+\infty$
for any $n\in \N$.
Since $\suplim_{x\rightarrow 0+}\frac{(u')^{-1}(\lambda x)}{(u')^{-1}(x)}$
is non-increasing in $\lambda$,
$\suplim_{x\rightarrow 0}\frac{(u')^{-1}(\lambda x)}{(u')^{-1}(x)}<+\infty$ for any $0<\lambda<1$.

Similarly, one can prove that
$\suplim_{x\rightarrow +\infty}\frac{u'(kx)}{u'(x)}<1$ for any $k>1$ if and only if
$\exists\; \bar k>1$ such that $\suplim_{x\rightarrow +\infty}\frac{u'(\bar k x)}{u'(x)}<1$.

Now, suppose $L=\suplim_{x\rightarrow 0+}\frac{(u')^{-1}(\frac{1}{2}x)}{(u')^{-1}(x)}<+\infty$
(notice that $L\ge 1$). Then
there exists $\delta>0$ such that for any $x\in (0, \delta]$,
\begin{eqnarray*}
&&\frac{(u')^{-1}(\frac{1}{2}x)}{(u')^{-1}(x)}\le 2L\\
&\Rightarrow& \frac{1}{2}x \ge u'(2L (u')^{-1}(x))\\
&\Rightarrow& \frac{1}{2} \ge \frac{u'(2L (u')^{-1}(x))}{u'((u')^{-1}(x))}\\
&\Rightarrow & \frac{u'(2Ly)}{u'(y)}\le \frac{1}{2}, \qquad \forall\; y\ge (u')^{-1}(\delta)\\
&\Rightarrow & \suplim_{x\rightarrow +\infty}\frac{u'(2L x)}{u'(x)}\le \frac{1}{2}.
\end{eqnarray*}
Therefore $\suplim_{x\rightarrow +\infty}\frac{u'(k x)}{u'(x)}< 1$ for any $k>1$.

The proof for the other direction is similar. \eof


Recall that we have defined $f(\lambda)=E[(u')^{-1}(\lambda \xi)\xi]$ and
$\lambda_0=\inf\{\lambda>0: f(\lambda)<+\infty\}.$

\begin{prop}\label{conslambda}
  Suppose one of the following conditions is satisfied:
  \begin{itemize}
    \item [{\rm (i)}]$\sublim_{x\rightarrow +\infty} R(x)>0$.
    \item [{\rm (ii)}]$\suplim_{x\rightarrow +\infty}\frac{u'(k x)}{u'(x)}< 1$ for some $k>1$.
    \item [{\rm (iii)}]$\suplim_{x\rightarrow 0}\frac{(u')^{-1}(\lambda x)}{(u')^{-1}(x)}<+\infty$
    for some $\lambda\in (0,1)$.
  \end{itemize}
Then the Lagrange multiplier exists for any $a>0$ if and only if
  $\lambda_0<+\infty$.
\end{prop}
\pf The necessity is obvious.
To prove the sufficiency,
note that if $\lambda_0<+\infty$, then there exists $\lambda_1>0$ such that $f(\lambda_1)<+\infty$,
which by the monotonicity of $f(\cdot)$
further implies that
$f(\lambda)<+\infty$
$\forall \lambda>\lambda_1$.
For any $\lambda\in (0, \lambda_1]$, denote $k=\lambda/\lambda_1\in (0,1]$.

Since one of the three given conditions is satisfied,
by Lemmas \ref{lbra} and \ref{u'inv} it must have
$1\le L=\suplim_{x\rightarrow 0}\frac{(u')^{-1}(kx)}{(u')^{-1}(x)}<+\infty$.
Hence there exists $\delta>0$ such that $\frac{(u')^{-1}(kx)}{(u')^{-1}(x)}<2L$ for any
$x\in (0, \lambda_1\delta]$. Now, for any $\lambda>0$,
\begin{eqnarray*}
  E[(u')^{-1}(\lambda \xi)\xi\id_{\xi\le \delta}]
  &=&E\left[\frac{(u')^{-1}(\lambda \xi)}{(u')^{-1}(\lambda_1\xi)}(u')^{-1}(\lambda_1\xi)
  \xi\id_{\xi\le \delta}\right]\\
  &\le &2LE[(u')^{-1}(\lambda_1\xi)\xi\id_{\xi\le \delta}]\\
  &\le &2Lf(\lambda_1),\\
  E[(u')^{-1}(\lambda \xi)\xi\id_{\xi>\delta}]&=&\frac{1}{\lambda}E[(u')^{-1}(\lambda \xi)
  (\lambda \xi)\id_{\xi>\delta}]\\
  &\le& \frac{1}{\lambda}E[u((u')^{-1}(\lambda \xi))\id_{\xi>\delta}]\\
  &\le&\frac{1}{\lambda}u((u')^{-1}(\lambda \delta)).
\end{eqnarray*}
Hence,
\begin{eqnarray*}
  f(\lambda)&=&E[(u')^{-1}(\lambda \xi)\xi]\\
  &=&E[(u')^{-1}(\lambda \xi)\xi\id_{\xi\le \delta}]+E[(u')^{-1}(\lambda \xi)\xi\id_{\xi>\delta}]\\
  &\le&2Lf(\lambda_1)+\frac{1}{\lambda}u((u')^{-1}(\lambda \delta))\\
  &<&+\infty.
\end{eqnarray*}
This shows that in fact $\lambda_0=0$, and hence
the equation $f(\lambda)=a$ admits a positive solution $\lambda(a)$
for any $a>0$.\eof

\begin{remark}
{\rm The preceding proof also shows that under the condition of
Proposition \ref{conslambda}, the following claims are equivalent:
  \begin{itemize}
    \item [(i)] The Lagrange multiplier exists for any $a>0$.
    \item [(ii)] $\lambda_0<+\infty$.
    \item [(iii)] $\lambda_0=0$.
    \item [(iv)] $f(1)<+\infty$.
\item [(v)] $f(\lambda)<+\infty$ $\forall \lambda>0$.
  \end{itemize}}
\end{remark}

\begin{theo}\label{sovability}
Under the condition of Proposition \ref{conslambda},
Problem (\ref{maxut}) admits
a unique optimal solution for any $a>0$ if and only if $E[u((u')^{-1}(\xi))]<+\infty$.
\end{theo}
\pf It suffices to prove the sufficiency.
If $E[u((u')^{-1}(\xi))]<+\infty$, then
$  f(1)=E[(u')^{-1}(\xi)\xi]\le E[u((u')^{-1}(\xi))]<+\infty.$
Thus $\lambda_0=0$ and $a_0=f(\lambda_0+)=+\infty$. It follows from
Proposition \ref{gebeexist} then that
Problem (\ref{maxut}) admits
a unique optimal solution.
\eof

The conditions in the preceding theorem, $\sublim_{x\rightarrow +\infty} -\frac{xu^\dpm(x)}{u'(x)}\ge 0$ and
$E[u((u')^{-1}(\xi))]<+\infty$,
are very easy to verify. For example, a commonly used utility function is
$u(x)=x^{\alpha}$, $0<\alpha<1$. The two conditions are satisfied when
$\xi$ is lognormal.

\begin{remark}
{\rm Example \ref{nonattain} shows that the conclusion of
Theorem \ref{sovability} can be false in the absence of
its condition.}
\end{remark}

\begin{coro}
If $E[\xi^{-\alpha}]<+\infty$ $\forall \alpha\geq1$, then, under the condition of Proposition \ref{conslambda},
Problem (\ref{maxut}) admits
a unique optimal solution for any $a>0$.
\end{coro}
\pf
It suffices to prove that $E[u((u')^{-1}(\xi))]<+\infty$ holds automatically.
Under the condition of Proposition \ref{conslambda}, there
is $L\geq 2$ such that
$(u')^{-1}(x)<L(u')^{-1}(2x)$ $\forall x\in(0,1)$.
Denote $L_0=\sup_{x\in[\frac{1}{2},1]}(u')^{-1}(x)<+\infty$. For any $x\in(0,1)$,
find $n\in\N$ so that $\frac{1}{2}\le 2^{n}x<1$. Then
$(u')^{-1}(x)<L(u')^{-1}(2x)<L^2(u')^{-1}(2^2x)<
\cdots<L^n(u')^{-1}(2^nx)\le L^nL_0\le L^{-\log_2 x}L_0=x^{-\log_2 L}L_0$.
By virtue of the fact that $u'(+\infty)=0$, we may assume that
$u(x)\leq L_1 x$ $\forall x\ge (u')^{-1}(1)$. Therefor for any $x\in(0,1)$,
we have  $u((u')^{-1}(x))\leq L_1(u')^{-1}(x)<L_0L_1x^{-\log_2 L}$.
Finally, 
$ E[u((u')^{-1}(\xi))]\leq E[u((u')^{-1}(\xi))\mathbf{1}_{\xi<1}]+u((u')^{-1}(1))
\leq L_0L_1E[\xi^{-\log_2 L}]+u((u')^{-1}(1))<+\infty.
$
\eof

\begin{remark}
{\rm If $\xi$ is lognormal, then the assumption that $E[\xi^{-\alpha}]<+\infty$ $\forall \alpha\geq1$ holds automatically.
(In the context of portfolio selection with the prices of the underlying
stocks following geometric Brownian motion, $\xi$ is typically a lognormal
random variable -- under certain conditions of course; for details see
Remark 3.1 in Beliecki et. al (2005).)
On the other hand, this assumption could be
weakened to that $E[\xi^{-\alpha_0}]<+\infty$
for {\it certain} $\alpha_0$ (the value of which could be precisely given).
We leave the details to the interested readers.}
\end{remark}

Recall that in Section 4 we presented an example where Problem (\ref{maxut}) is ill-posed
even though the Lagrange multiplier exists for any $a>0$. The following
result shows that this will not occur for certain $\xi$.

Let $F(\cdot)$ be the probability distribution function of $\xi$.
In view of Theorem \ref{einf0}, we assume
${\rm essinf}\;\xi=0$,
which in turn ensures $F(x)>0$ $\forall x>0$.

\begin{theo}\label{extra}
If $\liminf_{x\to 0}\frac{xF'(x)}{F(x)}>0$,
and $E[(u')^{-1}(\lambda\xi)\xi]=a>0$ for some $\lambda>0$,
then Problem (\ref{maxut}) with parameter $a$ is well-posed and admits
a unique optimal solution.
\end{theo}
\pf Since $\liminf_{x\to 0}\frac{xF'(x)}{F(x)}>0$,
there exist $M>0$ and $K>0$ such that $\frac{xF'(x)}{F(x)}\ge \frac{1}{K}$
for any $0<x\le M$. Then
\begin{eqnarray*}
&&E[u((u')^{-1}(\lambda\xi))\mathbf{1}_{\xi<M}]\\
&=&\int_0^Mu((u')^{-1}(\lambda x))d F(x)\\
&=&\int_0^M\int_{M}^x du((u')^{-1}(\lambda y))d F(x)+\int_0^Mu((u')^{-1}(\lambda M))d F(x)\\
&=&\lambda\int_0^M\int_{M}^x yd[(u')^{-1}(\lambda y)]d F(x)+u((u')^{-1}(\lambda M))F(M)\\
&=&\lambda\int_0^M\left(x(u')^{-1}(\lambda x)-M(u')^{-1}(\lambda M)+
\int_x^{M} (u')^{-1}(\lambda y)dy\right)d F(x)\\
&&+u((u')^{-1}(\lambda M))F(M)\\
&=&\lambda\int_0^M x(u')^{-1}(\lambda x)d F(x)+
\lambda \int_0^M\int_x^{M} (u')^{-1}(\lambda y)dy d F(x)\\
&&+[u((u')^{-1}(\lambda M))-\lambda M(u')^{-1}(\lambda M)]F(M)\\
&=&\lambda \int_0^M x(u')^{-1}(\lambda x)d F(x)
+\lambda \int_0^{M}\int_0^{y} d F(x)(u')^{-1}(\lambda y)dy\\
&&+[u((u')^{-1}(\lambda M))-\lambda M(u')^{-1}(\lambda M)]F(M)\\
&=&\lambda \int_0^M x(u')^{-1}(\lambda x)d F(x)+\lambda \int_0^{M}F(y)(u')^{-1}(\lambda y)dy\\
&&+[u((u')^{-1}(\lambda M))-\lambda M(u')^{-1}(\lambda M)]F(M)\\
&\leq &\lambda\int_0^M x(u')^{-1}(\lambda x)d F(x)+K\lambda\int_0^{M}y F'(y)(u')^{-1}(\lambda y)dy\\
&&+[u((u')^{-1}(\lambda M))-\lambda M(u')^{-1}(\lambda M)]F(M)\\
&\leq &\lambda(1+K)a+[u((u')^{-1}(\lambda M))-\lambda M(u')^{-1}(\lambda M)]F(M)\\
&<&+\infty.
\end{eqnarray*}
Consequently,
\begin{eqnarray*}
&&E[u((u')^{-1}(\lambda \xi))]\\
&=&E[u((u')^{-1}(\lambda\xi))\mathbf{1}_{\xi<M}]+E[u((u')^{-1}(\lambda\xi))\mathbf{1}_{\xi\ge M}]\\
&\le &E[u((u')^{-1}(\lambda\xi))\mathbf{1}_{\xi<M}]+u((u')^{-1}(\lambda M))\\
&<&+\infty.
\end{eqnarray*}
The desired result follows then from Theorem \ref{generallag}.
\eof

\begin{remark}
{\rm The condition $\liminf_{x\to 0}\frac{xF'(x)}{F(x)}>0$
implicitly requires that $F(\cdot)$ be differentiable in the neighborhood of
0. Notice that this requirement is purely technical so as to make the result
neater. Once could replace the condition $\liminf_{x\to 0}\frac{xF'(x)}{F(x)}>0$
by a weaker one without
having to assume the differentiability of $F(\cdot)$ (as hinted by the
preceding proof -- the details are left to the interested reader). On the
other hand, the condition
is satisfied if $\xi$ is lognormal.}
\end{remark}

Combining Theorem \ref{generallag} and Theorem \ref{extra}, we have immediately

\begin{coro}\label{coro3}
Suppose $\liminf_{x\to 0}\frac{xF'(x)}{F(x)}>0$.
Then Problem (\ref{maxut}) with parameter $a>0$ admits an optimal solution
if and only if the Lagrange multiplier $\lambda$ exists corresponding to
$a$, in which case the unique optimal solution
is $X^*=(u')^{-1}(\lambda\xi)$.
\end{coro}

The following synthesized result gives easily verifiable conditions under which
Problem (\ref{maxut}) is completely solved.

\begin{theo}\label{grand} We have the following conclusions.
\begin{itemize}
  \item[{\rm (i)}] If $\sublim_{x\rightarrow +\infty} \left(-\frac{xu^\dpm(x)}{u'(x)}\right)>0$, then the following statements are equivalent:
\begin{itemize}
  \item[{\rm (ia)}] Problem (\ref{maxut}) is well-posed for any $a>0$.
\item[{\rm (ib)}] Problem (\ref{maxut}) admits a unique optimal
solution.
\item[{\rm (ic)}] $E[u((u')^{-1}(\xi))]<+\infty$.
\item[{\rm (id)}] $\exists$ $\lambda>0$ such that $E[u((u')^{-1}(\lambda\xi))]<+\infty$.
\end{itemize}
Moreover, when one of {\rm (ia)}--{\rm (id)} holds
the optimal
solution to (\ref{maxut}) with parameter $a>0$
is $X^*=(u')^{-1}(\lambda(a)\xi)$, where $\lambda(a)$
is the Lagrange multiplier corresponding to $a$.
\item[{\rm (ii)}] If $\limsup_{x\to 0}\left(-\frac{xF'(x)}{F(x)}\right)<0$,
then Problem (\ref{maxut}) is well-posed for any $a>0$ if and only
if $E[(u')^{-1}(\lambda\xi)\xi]<+\infty$ for some $\lambda>0$, in which case
there exists $0<a_0\leq +\infty$ so that (\ref{maxut})
admits a unique optimal
solution $X^*=(u')^{-1}(\lambda(a)\xi)$ for any $a>0$ (if $a_0=+\infty$)
or for any $0<a
\leq a_0$ (if $a_0<+\infty$).
\end{itemize}
\end{theo}

\pf (i) If (\ref{maxut}) is well-posed for any $a>0$, then
Theorem \ref{lambda0infty}
yields that $f(\lambda_0)<+\infty$ for some $\lambda_0>0$. It follows
from Proposition \ref{conslambda} and Theorem \ref{generallag} that (\ref{maxut}) admits a unique
optimal solution for any $a>0$. The desired equivalence is then a consequence
of Theorem \ref{sovability} and Theorem \ref{generallag}.

(ii) The first conclusion (``if and only if'')
follows from Theorems \ref{lambda0infty} and \ref{extra}.
For the second conclusion, let $\lambda_0=\inf\{\lambda>0: f(\lambda)<+\infty\}<+\infty$ and $a_0=f(\lambda_0+)$. Then the Lagrange multiplier exists for
any $a>0$ (if $a_0=+\infty$) or for any $0<a
\leq a_0$ (if $a_0<+\infty$), and
Corollary \ref{coro3} completes the proof.
\eof

\begin{remark}
{\rm Portfolio selection is essentially an endeavor that an investor,
given a market (represented by $\xi$ or its
distribution function $F(\cdot)$), tries to make the best out of his initial wealth (namely $a$) taking advantage of the availability of the market,
where the ``best'' is measured by her preference (i.e.
the utility function $u(\cdot)$). We have shown that these entities, namely
$F(\cdot)$, $a$, and $u(\cdot)$, must coordinate well, otherwise one
may end up with a wrong model. The assumptions stipulated
in Theorem \ref{grand} tell precisely how this well-coordination
can be translated into mathematical conditions.}
\end{remark}

\section{Concluding Remarks}

The stochastic optimization problem studied in this paper, though
interesting in its own right, has profound
applications in financial asset allocation among others.
It is demonstrated that many assumptions that have been taken for granted,
such as the well-posedness of the problem, existence of the Lagrange
multiplier, and existence of an optimal solution, may be invalid
in the first place.  In particular, the issue of well-posedness is
equally important, if not more important, than that of finding an optimal
solution from a modeling point of view.
Attainability of
optimal solutions is another important matter: if an optimal solution
is not attainable, as is the case with Example \ref{nonattain}, then
one has to resort to finding an asymptotically optimal solution.
Mathematically, both the ill-posedness and the non-attainability
are symptomized by the non-existence of the Lagrange multiplier, as
analyzed in details in this paper.

It is worth noting that the results of this paper have been utilized in
solving a sub-problem
of the continuous-time behavioral portfolio selection model Jin and Zhou (2006), where
the ill-posedness is more a rule than an exception.


\newpage


\begin{thebibliography}{99}

{\hsc T.R. Bielecki, H. Jin, S.R. Pliska and X.Y. Zhou} (2005):
{Continuous-time mean--variance portfolio selection with bankruptcy
prohibition}, {\it Math. Finance} 15, pp. 213-244.\\

{\hsc J. Cvitanic and I. Karatzas}(1992), {Convex duality in constrained
portfolio optimization}, {\it Ann. Appl. Probab.} 2, pp. 767-818.\\

{\hsc I. Karatzas} (1997), {\it Lectures on the Mathematics of Finance},
American Mathematical Society.\\

{\hsc I. Karatzas and S.E. Shreve} (1998), {\it Methods of Mathematical Finance},
Springer-Verlag, New York.\\

{\hsc R. Korn} (1997), {\it Optimal Portfolios},
World Scientific, Singapore.\\

{\hsc R. Korn and H. Kraft} (2004), {On the stability of continuous-time portfolio problems
with stochastic opportunity set}, {\it Math. Finance} 14, pp. 403-414.\\

{\hsc H. Jin and X.Y. Zhou} (2006),
{Behavioral portfolio selection in continuous time}, Working paper,
Department of Systems Engineering and Engineering
Management, The Chinese University of Hong Kong.\\

\end{thebibliography}
\end{document}